\documentclass[aps,prab,groupedaddress,floats,reprint]{revtex4-2}

\usepackage[english] {babel}
\usepackage{amsfonts,graphicx,soul,natbib,mathrsfs}

\usepackage[colorlinks=true]{hyperref}

\usepackage{epstopdf}
\usepackage{xcolor}
\usepackage{soul}
\usepackage{physics}
\usepackage{graphics}
\usepackage{comment}
\newcommand{\p}{\partial}
\renewcommand{\vec}[1]{\textnormal{\boldmath$#1$}}

\usepackage{dcolumn}
\usepackage{bm}
\begin{document}

\title{Effects of the transversely nonuniform plasma density in a blowout regime of a plasma wakefield accelerator}

\author{S. S. Baturin}%
\email{s.s.baturin@gmail.com}%
\affiliation{School of Physics and Engineering,
ITMO University, St. Petersburg, Russia 197101}%

\date{\today}
\begin{abstract}
We present an analytical study on the effects of the transverse plasma gradient in the blowout regime of a plasma wakefield accelerator (PWA). The analysis departs from a simple ballistic model of plasma electrons that allows us to derive a complete analytic solution for the pseudopotential and, consequently, for the wakefield. We demonstrate that the transverse plasma gradient modifies the bubble shape and affects the wakefield. Namely, the dipole plasma gradient results in a dipole component of the wakefield. Analysis suggests that, despite the asymmetry, the instability due to the fixed transverse plasma gradient is unlikely, as the total wakefield has a single stable point inside the bubble. The only effect that occurs is the shift of the electromagnetic center. We point out that random fluctuation of the transverse plasma gradient could become an issue.     
\end{abstract}

\maketitle


\section{Introduction}


Beam breakup (BBU) and emittance degradation are one of the main challenges on the way to the high luminosity collider based on a wakefield accelerator technology. \cite{Jing,Sasha2,Sasha3, Hosing1,BBU2,BBU3}. Among others, \cite{Jing,Sasha2,Sasha3} plasma-based wakefield accelerators (PWA) are considered the most promising candidate for the particle accelerators of the future  \cite{SNM3,LK,alegro-2019-a,Adli,Sasha1,Jamie1} as it possesses intrinsic focusing mechanism provided by the ions and extreme accelerating gradients at the same time \cite{Rosen1,Rosen2}. In particular, in a so-called bubble regime, when plasma electrons are expelled, due to the high intensity of either laser or electron driver longitudinal (accelerating) electric field could be $\sim$ 50 GV/m \cite{WL,Blumenfeld:2007tz}. Accelerating gradient is tightly connected with the transverse wakefield by the means of the Panofsky-Wenzel theorem \cite{Panofsky:1956}. When the longitudinal wake is axisymmetric, the transverse wake is zero, but even a tiny asymmetry is enough to seed the instability, that is caused by the transverse wake. The latter is connected with the fact that projected transverse beam emittance grows exponentially due to the BBU \cite{Chao}. Moreover, even a short noise (a small random force or random asymmetry in the beam distribution) triggers the BBU.               

Despite such a strong limitation, it tums out that BBU could still be controlled. Many mechanisms and approaches, including ion motion \cite{Timon2}, BNS damping \cite{Lehe,Timon1}, and other methods of instability suppression have been investigated extensively. It was demonstrated that BBU is not just suppressed but eliminated in a certain parameter range.   

In the present study, we investigate how the local transverse asymmetry of the plasma gradient affects the wake and analyze these results from a beam dynamics perspective.The most promising approach to the analytic (or semi-analytic) description of the bubble regime is the Lu model \cite{Lu, Mori}, unfortunately, it does not account for the transverse plasma gradient yet and works best behind the driver. Previous studies \cite{Puhov2016,Puhov2017} indicate that transverse plasma gradient (axisymmetric in the considered cases) modifies the wakefield and thus it must be taken into account. It turns out that the most promising high repetition PWA technology is the transverse flowing supersonic gas jet \cite{FLASHfwd}, as heat and transient effects in plasma become an issue \cite{Mike,DArcy:2022aa}. The drawback of this technology is built-in electron and ion density gradients \cite{PPG1,PPG2}. As a consequence, such a scheme might be vulnerable to the BBU. A recent study \cite{mikeprab} indicates that the transverse plasma gradient of a supersonic gas jet indeed results in a modified dynamics and echoes some conclusions of the present paper.    

The analysis is based on the ballistic model of the plasma electrons introduced in Ref.\cite{Stupakov:2016,Stupakov:2018}. To an extent, present calculations echo the analysis of the Ref.\cite{FlatPRAB}, where a flat bubble formation was investigated. In contrast to the Ref.\cite{FlatPRAB}, where it was demonstrated that the wakefield is insensitive to the transverse plasma gradient, we show that in a commonly considered round bubble, transverse plasma gradient results in a transverse wake. Within the considered approximation we analyze plasma flow, derive an analytic expression for the pseudopotential, and provide a discussion of the possible consequences.

The paper is organized as follows. In Sec.\ref{sec:BE}, for the sake of convenience, we reproduce basic formulas and equations provided in Refs.\cite{Stupakov:2016,Stupakov:2018,FlatPRAB} that we will use throughout the paper. In Sec.\ref{sec:SW} we derive the expression for the electromagnetic shock wave produced by the driver, and in Sec.\ref{sec:BBS} we present an expression that describes the bubble shape within ballistic approximation. In Sec.\ref{sec:pld} plasma electron density is derived, and the outcome of this section is then used in Sec.\ref{sec:PPT} to get to the main result of the paper - the expression for the pseudopotential.


\section{Basic equations \label{sec:BE}}


We utilize the general idea of the model introduced in Ref.\cite{Mora1997} and start from the set of equations derived in Ref.\cite{Stupakov:2018}.  We use the same convention as the Ref.\cite{Stupakov:2018} and we use dimensionless variables: time is normalized to $\omega_p^{-1}$, length to $k_p^{-1}$, 
velocities to the speed of light $c$, and momenta to $mc$. We also normalize fields to $mc\omega_p/e$, forces to $mc\omega_p$, potentials to $mc^2/e$, the charge density to $n_0 e$, the plasma density to $n_0$, and the current density to $en_0c$. With $e$ being the elementary charge, $e>0$.

The equations of motion for the plasma electrons could be written as \cite{Mora1997}
\begin{align}
\label{eq:ple}
\frac{d\mathbf{p}}{dt}=\mathbf{\nabla} \phi-\mathbf{v}\cross\mathbf{B},~\frac{d\mathbf{r}}{dt}=\frac{\mathbf{p}}{\gamma}.
\end{align}
Here $\mathbf{p}$ is the momentum of the plasma electrons, $\gamma=\sqrt{1+p^2}$ is the relativistic gamma factor of the plasma electrons, $\mathbf{v}=\mathbf{p}/\gamma$ is the velocity and $\phi$ is the electric potential. Additionally, we introduce pseudopotential $\psi=\phi-A_z$ ($A_z$ is the $z$ component of the vector potential) that defines that wakefield as
\begin{align}
\label{eq:lor_main}
E_z=\frac{\partial \psi}{\partial \xi}, ~~~\mathbf{F}_{\perp}=-\mathbf{\nabla}_{\perp}\psi,
\end{align}
and $\mathbf{\nabla}=(\partial_x,\partial_y,-\partial_\xi)$.
Here $\mathbf{F}_{\perp}$ is the transverse part of the Lorentz force  per unit charge of the test particle and $\xi=t-z$.

In the quasistatic approximation, when the Hamiltonian that corresponds to the Eqs.\eqref{eq:ple} depends on $z$ and $t$ only in combination $\xi=t-z$ there exists the following integral of motion \cite{Mora1997,Stupakov:2018}
\begin{align}
\gamma-p_z-\psi=1,
\end{align}
as a consequence we have 
\begin{align}
1-v_z=\frac{1+\psi}{\gamma}.
\end{align}
In a quasi-static picture, it is convenient to replace the derivative by time $t$ with the derivative by $\xi$.
We use the fact that
\begin{align}
\frac{d\xi}{dt}=1-v_z,
\end{align}
consequently for an arbitrary function $f(\xi)$ we have.
\begin{align}
\label{eq:con}
\frac{df}{dt}=\frac{df}{d\xi}\frac{d\xi}{dt}=(1-v_z)\frac{df}{d\xi}=\frac{1+\psi}{\gamma} \frac{df}{d\xi}.
\end{align}
Since in the quasi-static picture momentum of the plasma electron is a function of $\xi$ Eqs.\eqref{eq:ple} with Eq.\eqref{eq:con} are reduced to
\begin{align}
\frac{d\mathbf{p}_\perp}{d\xi}=\frac{\gamma}{1+\psi}\left[\mathbf{\nabla}_\perp \psi+\mathbf{\hat{z}}\cross \mathbf{B}_\perp\right]-\frac{B_z}{1+\psi}\mathbf{p}_\perp\cross\mathbf{\hat{z}}.
\end{align}
Equation for the pseudopotential reads textcolor{red}{(see Ref.\cite{Stupakov:2018})}
\begin{align}
\label{eq:WP}
\Delta_\perp\psi=(1-v_z)n_e-n_i(x),
\end{align}
 here $n_e$ is the plasma electron density and $n_i(x)$ is the ion density that depends on $x$. In what follows, we will assume that
 \begin{align}
     n_i(x)
     =
     1+gx,
 \end{align}
 with $g\ll 1$. The assumption above is dictated by the natural gradient of the gas density in a gas jet. Fast ionization and immediate interaction make both densities equal if one assumes that jet gradient is dominating over others.

 Equations for the magnetic field are 
 \begin{align}
& \Delta_\perp B_z=\mathbf{\hat{z}}\cdot(\mathbf{\nabla_\perp}\cross n_e\mathbf{v}_\perp), \\
 &\Delta_\perp \mathbf{B}_\perp=-\mathbf{\hat{z}}\cross\mathbf{\nabla}_\perp n_e v_z-\mathbf{\hat{z}}\cross \partial_\xi n_e \mathbf{v}_\perp.
 \end{align}

The continuity equation reads
\begin{align}\label{eq:cont_equation}
    \p_\xi[n_e(1-v_z)]
    +
    \nabla_\perp
    \cdot
    n_e\vec v_\perp
    =
    0
    .
\end{align}

%
\section{Shock wave \label{sec:SW}}
%

To calculate the field distribution that is produced by the point driver that travels through plasma, we follow  Refs.\cite{Stupakov:2016,Stupakov:2018}. Namely, we assume that driver fields are localized in an infinitesimally thin layer, i.e.  
have a delta-function discontinuity

\begin{align}
    \mathbf{E}_\perp&=\mathbf{D} \delta(\xi), \nonumber \\
    \mathbf{B}_\perp&=\mathbf{\hat{z}}\times \mathbf{D} \delta(\xi). 
\end{align}

The transverse profile of these fields is defined by the 2D vector $\mathbf{D}$.

To solve for the shock wave at $\xi=0$, we assume that the plasma density in front of the moving driver has a linear gradient
    \begin{align}\label{14}
        n_{0}
        =
        1 + gx
        ,
    \end{align}
where the uniform part of the density is 1, $g$ is a constant, and $x$ is the transverse coordinate. We assume a small gradient,
    \begin{align}
        g\ll 1
        ,
    \end{align}
and use the perturbation theory.

We consider equation for the vector $\vec D$ that according to Ref.\cite{Stupakov:2018} reads
\begin{align}
\label{eq:Dg}
\Delta_\perp \vec D =\frac{n_0}{\gamma_0} \vec D.
\end{align}
If we split $\vec D$ into $r$ and $\phi$ component then Eq.\eqref{eq:Dg} could be written in expanded form as
\begin{align}
\label{eq:Dg2}
&\Delta_\perp D_r-\frac{D_r}{r^2}-\frac{2}{r^2}\frac{\partial D_\phi}{\partial \phi} =\frac{n_0}{\gamma_0} D_r, \nonumber \\
&\Delta_\perp D_\phi-\frac{D_\phi}{r^2}+\frac{2}{r^2}\frac{\partial D_r}{\partial \phi} =\frac{n_0}{\gamma_0} D_\phi,
\end{align}
with the Laplace operator given by
\begin{align}
   \Delta_\perp=\frac{1}{r}\frac{\partial}{\partial r}\left[r \frac{\partial }{\partial r}\right]+\frac{1}{r^2}\frac{\partial^2 }{\partial \phi^2}.
\end{align}
If we assume that $n_0$ is a constant then due to the axial symmetry we have $\vec D=D_r(r)\vec r$ and $D_\phi=0$.
This immediately results in the equation for $D(r)$ in a form
\begin{align}
\label{eq:Dr2}
&\frac{1}{r}\frac{\partial}{\partial r}\left[r \frac{\partial D_r}{\partial r}\right]-\frac{1}{r^2}D_r =\frac{n_0}{\gamma_0} D_r.
\end{align}
With the unmodified plasma density, initial gamma set to unity ($n_0=1$, $\gamma_0=1$) and boundary conditions $D_r(\infty)=0$, $D_r(r\to 0)=2\nu/r$ we have
\begin{align}
\label{eq:D01}
    D_r(r,\phi)=2\nu K_1(r).
\end{align}
Next, we consider $n_0$ as given by Eq.\eqref{14}. We apply perturbation theory and seek a solution of the Eq.\eqref{eq:Dg2} in a form
\begin{align}
\label{eq:ppr221}
    &D_r=D_r^{(0)}+D_r^{(1)}, \\
    &D_\phi=D_\phi^{(0)}+D_\phi^{(1)}. \nonumber
\end{align}
With the $D_r^{(0)}=2\nu K_1(r)$, $D_\phi^{(0)}=0$ and $D_{r,\phi}^{(1)}\sim g$ - small corrections. Substituting Eqs.\eqref{eq:ppr221} and Eq.\eqref{14} into the Eqs.\eqref{eq:Dg2}, equating terms of the same order and accounting for $x=r\cos(\phi)$ we arrive at a set of equations for corrections in the form
\begin{align}
\label{eq:DcorP22}
&\Delta_\perp D^{(1)}_r-\frac{D^{(1)}_r}{r^2}-\frac{2}{r^2}\frac{\partial D^{(1)}_\phi}{\partial \phi} =D_r^{(1)}+2\nu g K_1(r)r\cos(\phi), \nonumber \\
&\Delta_\perp D^{(1)}_\phi-\frac{D^{(1)}_\phi}{r^2}+\frac{2}{r^2}\frac{\partial D^{(1)}_r}{\partial \phi} =D^{(1)}_\phi.
\end{align}
Next we decompose $D_r^{(1)}$ and $D_\phi^{(1)}$ in a Fourier series
\begin{align}
\label{eq:FD}
    &D_r^{(1)}(r,\phi)=\sum \tilde D^r_n (r) \cos(n \phi), \\
    &D_\phi^{(1)}(r,\phi)=\sum \tilde D^\phi_n (r) \sin(n \phi),
\end{align}
and substitute this decomposition into Eqs.\eqref{eq:DcorP22}. Equating amplitudes of corresponding cosines and sines we have
\begin{align}
\label{eq:DcorFF22}
&\frac{1}{r}\frac{\partial}{\partial r}\left[r \frac{\partial D^r_1}{\partial r}\right]-\frac{2}{r^2} D^r_1-\frac{2}{r^2}D^\phi_1=D^r_1+2\nu g K_1(r)r, \nonumber \\
&\frac{1}{r}\frac{\partial}{\partial r}\left[r \frac{\partial D^\phi_1}{\partial r}\right]-\frac{2}{r^2}D^\phi_1-\frac{2}{r^2} D^r_1 =D^\phi_1.
\end{align}
Next we introduce new functions $\Psi=D_1^r+D_1^\phi$ and $\Phi=D_1^r-D_1^\phi$. Adding and subtracting Eqs.\eqref{eq:DcorFF22} we get
\begin{align}
\label{eq:DcorFF2m}
&\frac{1}{r}\frac{\partial}{\partial r}\left[r \frac{\partial \Psi}{\partial r}\right]-\frac{4}{r^2}\Psi=\Psi+2\nu g K_1(r)r, \nonumber \\
&\frac{1}{r}\frac{\partial}{\partial r}\left[r \frac{\partial \Phi}{\partial r}\right]=\Phi+2\nu g K_1(r)r.
\end{align}
A general solution to the equations above is zero as none of the functions fulfill boundary conditions $\Psi(\infty)\to 0$, $\Phi(\infty\to 0)$ and $\Psi(0)<\infty$, $\Phi(0)<\infty$. 
A specific solution on the other hand does not vanish and could be found via Hankel transformation. Forward and inverse Hankel transformation of the order $n$ of some function $f(r)$ are given by
\begin{align}
\label{eq:FH}
    &\hat f(k)=\mathcal{H}_n\left[f(r)\right]\equiv\int\limits_{0}^{\infty}r f(r) J_n(k r) dr, \\
\label{eq:RH}
    &f(r)=\mathcal{H}_n^{-1}\left[\hat f(k)\right]\equiv\int\limits_{0}^{\infty}k \hat f(k) J_n(k r) dk.
\end{align}
Here $J_n(k r)$ is the Bessel function of the first kind of the order $n$.

We apply $ \mathcal{H}_2$ to the fist equation and $ \mathcal{H}_0$ to the second equation and get
\begin{align}
&\tilde \Psi=-4\nu g\frac{k^2}{(1+k^2)^2}, \nonumber \\
&\tilde \Phi=-4\nu g\frac{1}{(1+k^2)^2}.
\end{align}
Inverse Hankel transform gives
\begin{align}
&\Psi=-4\nu g\mathcal{H}_2^{-1}\left[\frac{k^2}{(1+k^2)^3}\right]=-\frac{\nu g}{2}r^2K_0(r), \nonumber \\
&\Phi=-4\nu g\mathcal{H}_0^{-1}\left[\frac{1}{(1+k^2)^3}\right]=-\frac{\nu g}{2}r^2 K_2(r).
\end{align}
$D^r_1$ and $D^\phi_1$ are recovered as
\begin{align}
    &D^r_1=\frac{\Psi+\Phi}{2}, \\
    &D^\phi_1=\frac{\Psi-\Phi}{2},
\end{align}
and reads
\begin{align}
    &D^r_1=-\frac{\nu g}{4}r^2\left[K_0(r)+K_2(r) \right],\\
    &D^\phi_1=-\frac{\nu g}{4}r^2\left[K_0(r)-K_2(r)\right].
\end{align}
Finally, first order corrections could be written as
\begin{align}
\label{eq:Dcor2}
    &D_r^{(1)}=-\frac{\nu g}{4}r^2\left[K_0(r)+K_2(r) \right] \cos(\phi),\\
    &D_\phi^{(1)}=-\frac{\nu g}{4}r^2\left[K_0(r)-K_2(r)\right] \sin(\phi).
\end{align}


\section{Shape modification of the plasma bubble \label{sec:BBS}}

 
We neglect the effect of the plasma self-fields on the trajectories of the plasma electrons. This is a “ballistic” regime of plasma motion introduced in Ref.\cite{Stupakov:2016}; it assumes that the plasma electrons are moving with constant velocities.

We assume plasma electrons to be non-relativistic and $\nu \ll 1$ as well as  $\nu\ll r < 1$. With help of Eq.\eqref{eq:D01} and Eq.\eqref{eq:Dcor2} 
we may write equations of motion for the plasma electrons as  
\begin{align}
\label{eq:trvel}
    &\frac{dx}{d\xi}\approx\frac{2\nu}{r_0}\cos(\phi_0)-g\frac{\nu}{2}, \nonumber \\
    &\frac{dy}{d\xi}\approx\frac{2\nu}{r_0}\sin(\phi_0).
\end{align}
Solution to the equations above gives electron trajectories
\begin{align}
\label{eq:trjf}
    &x=r_0\cos{\phi_0}+2\nu \xi \left[\frac{\cos(\phi_0)}{r_0}-\frac{g}{4}\right], \nonumber \\
    &y=r_0\sin{\phi_0}+2\nu \xi \frac{\sin(\phi_0)}{r_0}.
\end{align}
From Eqs.\eqref{eq:trjf} one can deduce (see Appendix \ref{app:envs}) an equation of an envelope surface that defines the boundary of the bubble in the ballistic approximation in a form
\begin{align}\label{eq:103}
    \left(
    x+\frac{1}{2} \nu\xi g
    \right)^2
    +
    y^2
    =
    8\nu\xi
    .
\end{align}
It could be seen from the equation above that the circular cross sections of the bubble in $x,y$ plane are shifted in the $x$ direction by the distance $\nu \xi g/2$ that is linearly increasing with $\xi$.

We introduce the coordinates  $\tilde x = x+g\nu\xi /2$, $\tilde y=y$ and $\tilde{r} = \sqrt{\tilde x^2+\tilde{y}^2}$ and the angle $\tilde{\phi} = \arccos{\tilde{x}/ \tilde{r}}$. Then Eqs.\eqref{eq:trjf} can be written as
\begin{align}\label{eq:103a}
    \tilde{r}
    =
    r_0 + 2\nu \frac{\xi}{r_0},
    \qquad
    \tilde{\phi}
    =
    \phi_0
    .
\end{align}

Condition $d \tilde r/ d r_0=0$ gives equation for the bubble boundary 
\begin{align}
\label{eq:rb_sm}
    \tilde r_b(\xi)= 2\sqrt{2\nu \xi},
\end{align}
in full agreement with Ref.\cite{Stupakov:2016}.
Switching back to the $(x,y)$ we arrive at the final expression for the bubble boundary in the form
\begin{align}
  r_b(\xi,\phi)=2\sqrt{2\nu \xi}-g\frac{\nu \xi}{2}\cos\left( \phi \right).
\end{align}

\begin{figure}[t]
	\centering
	\includegraphics[width=1.\linewidth]{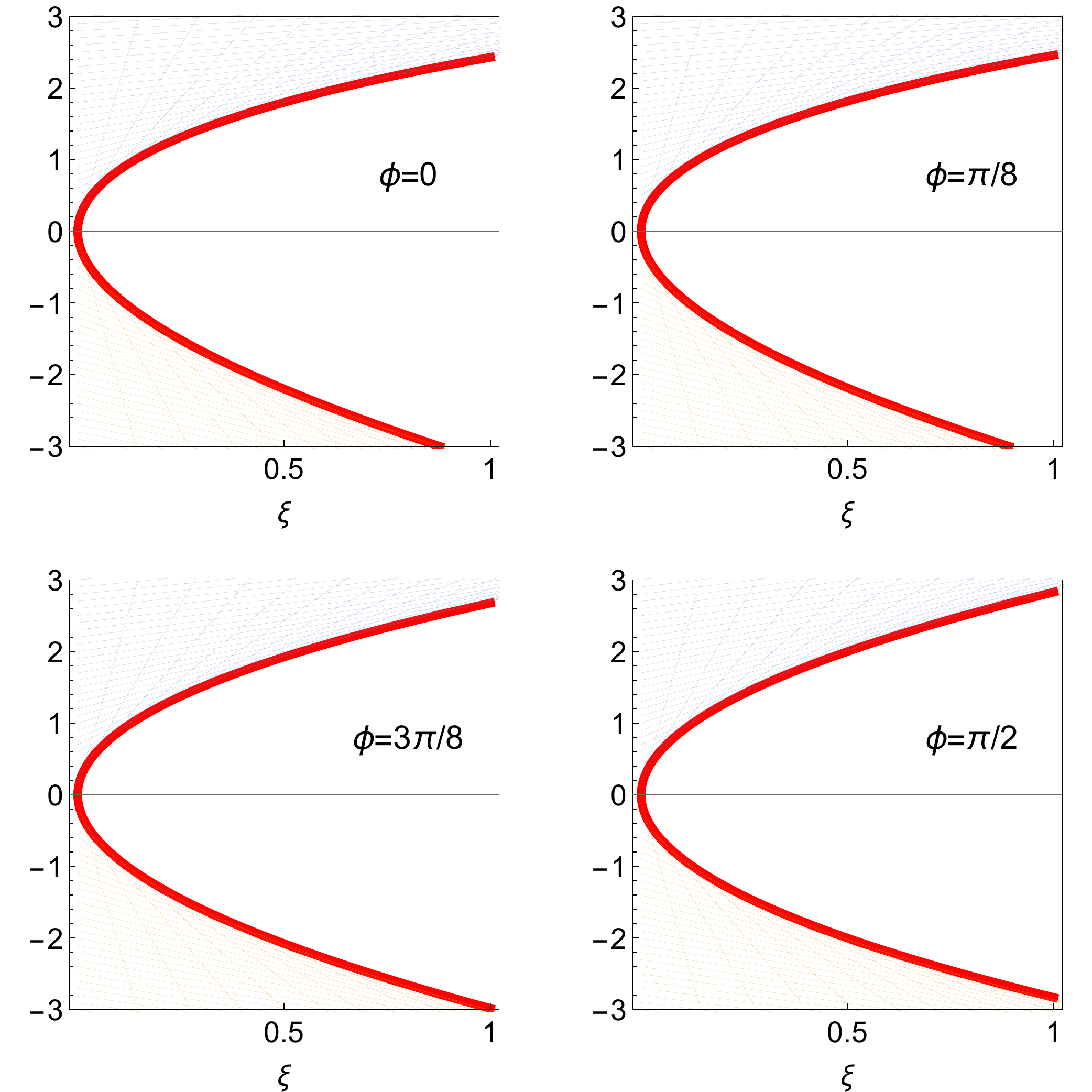}
	\caption{\label{fig:PSw} Longitudinal cuts of the plasma flow in the ballistic approximation for the case of $\nu=1$ and $g=0.8$. The numbers are far from realistic and are chosen to emphasise the effect visually. The vertical axis is the vertical coordinate of the plasma electrons in the corresponding cutting plane. The bubble is "bended" towards lower plasma density.}
\end{figure}

\section{Axial symmetry of the continuity equation in the ballistic approximation}

First, we refer to the continuity equation \eqref{eq:cont_equation} that in the ballistics approximation takes the form
\begin{align}\label{eq:cont_equation_bal}
    \p_\xi n_e +\nabla_\perp\cdot n_e\vec v_\perp = 0.
\end{align}
Here as before $\vec v_\perp$ is the vector of the transverse velocity of the plasma electrons that according to the Eq.\eqref{eq:trvel} reads
\begin{align}
\label{eq:vp}
   \vec v_\perp=\left(\frac{2\nu}{r_0}\cos(\phi_0)-g\frac{\nu}{2},\frac{2\nu}{r_0}\sin(\phi_0) \right)^{T}.
\end{align}
We notice that
\begin{align}
    &\frac{\partial}{\partial \tilde x}=\frac{\partial}{\partial x},\nonumber \\
    &\frac{\partial}{\partial \tilde y}=\frac{\partial}{\partial y}.
\end{align}

Next, if we assume that $n=n(\tilde{x},\tilde{y},\xi)$ then Eq.\eqref{eq:cont_equation_bal} should be modified as
\begin{align}
\label{eq:cont_equation_bal2}
    &\p_\xi n_e+g\frac{\nu}{2}\p_{\tilde{x}} n_e+v_x\p_{\tilde{x}}n_e+v_y\p_{\tilde{y}}n_e\nonumber\\&+n_e\p_{\tilde{x}}v_x+n_e\p_{\tilde{y}}v_y = 0.
\end{align}
Introducing $\mathbf{\tilde{v}}_\perp=\left(\frac{2\nu}{r_0}\cos(\phi_0),\frac{2\nu}{r_0}\sin(\phi_0) \right)^{T}$, $\tilde{\nabla}_\perp=(\p_{\tilde{x}},\p_{\tilde y})$ and noting that 
\begin{align}
    \p_{\tilde{x}}v_x=\p_{\tilde{x}}\tilde v_x,
\end{align}

we have 
\begin{align}
    \p_\xi n_e +\tilde{\nabla}_\perp\cdot n_e\mathbf{\tilde{v}}_\perp = 0.
\end{align}
We notice that in the new coordinates the plasma flow (in particular velocity field) has a cylindrical symmetry and thus continuity equation has a cylindrical symmetry as well and could be written as
\begin{align}
\label{eq:cont_equation_bal3}
    \p_\xi n_e +\frac{1}{\tilde r} \frac{\partial}{\partial \tilde r} \tilde r n_e \tilde v_\perp = 0.
\end{align}
with $\tilde v_\perp=\frac{2\nu}{r_0}$.

%
\section{Plasma density in the ballistic approximation \label{sec:pld}}
%

Immediately behind the driver, at $\xi=0^+$, the plasma density $n_0$ is given by Eq.\eqref{14}. 
If we assume that electron trajectories are known then from  the continuity of the plasma flow we conclude that 
\begin{align}
n(x,y,\xi) dS = n_0(x_0,y_0) dS_0, 
\end{align}
from which it follows that
\begin{align}
\label{eq:pldn}
n(x,y,\xi) = n_0(x_0,y_0) \frac{dS_0}{dS}. 
\end{align}
the ratio $dS/dS_0$ is calculated though the Jacobian
\begin{align}
    \frac{dS}{dS_0}=\left|
\begin{array}{cc}
 \frac{\partial x}{\partial x_0} & \frac{\partial x}{\partial y_0} \\
 \frac{\partial y}{\partial x_0} & \frac{\partial y}{\partial y_0} \\
\end{array}
\right|.
\end{align}
Near the bubble boundary, as far as maximum bubble radius $r_{bm}\sim \sqrt{\nu}$ and $\nu\ll 1$, we can use electron trajectories as defined by Eq.\eqref{eq:103a}. Accounting for the cylindrical symmetry in the $(\tilde x, \tilde y)$ coordinates we get 
    \begin{align}\label{eq:A.1}
    n(\tilde{r},\tilde{\phi},\xi)
    =&n_0(r_0,\phi_0)\frac{r_0dr_0}{\tilde{r}d \tilde{r}}= \nonumber \\
    &n_0(r_0,\phi_0) 
    \frac{r_0^3}{\tilde{r}|r_0^2-2\nu \xi|}
    ,
    \end{align}
where we have used Eq.\eqref{eq:103a} to calculate $d\tilde{r}/dr_0$. 

The initial radius $r_0$ in this equation should be expressed through $r$ and $\xi$ from Eq.~\eqref{eq:103a}:
    \begin{align}\label{eq:A.2}
    r_0^{\pm}
    =
    \frac{1}{2}
    \tilde{r}
    \pm
    \sqrt{
    \frac{1}{4}
    \tilde{r}^2
    -
    2\nu\xi}
    =
    \frac{1}{2}
    \tilde{r}
    \left(
    1
    \pm
    \sqrt{    1    -    t    }
    \right)
    ,
    \end{align}
where
    \begin{align}\label{eq:A.3}
    t
    =
    \frac{8\nu\xi}{\tilde{r}^2}
    =
    \frac{\tilde r_b^2}{\tilde{r}^2}
    <1
    ,
    \end{align}
with $\tilde r_b\equiv 2\sqrt{2\nu \xi}$ as given by Eq.\eqref{eq:rb_sm}.

Two solutions correspond to two trajectories that arrive from different initial radii $r_0,\phi_0$ to a given point $\tilde{r},\tilde{\phi},\xi$. One of these trajectories arrives before and the other one after it touches the envelope. Correspondingly, at a given $\xi,\tilde{r},\tilde{\phi}$, we need to sum the two densities for both trajectories.

Substituting~\eqref{eq:A.2} into~\eqref{eq:A.1} we obtain
    \begin{align}\label{eq:A.4}
    n_\pm(\tilde{r},\tilde{\phi},\xi)
    =
    \frac{1}{2}
    \frac{
    \left(
    1
    \pm
    \sqrt{    1    -    t    }
    \right)^3
    }{\big|
    \left(
    1
    \pm
    \sqrt{    1    -    t    }
    \right)^2
    -t\big|}
    (1+gr_0^\pm\cos\phi_0)
    ,
    \end{align}
where we have used Eq.~\eqref{14} for $n_0$. In this formula we have to express $r_0$ through $\tilde{r}$ using Eq.~\eqref{eq:A.2} and also use $\tilde{\phi} = \phi_0$.  For the total density, after some simplifications, we find
    \begin{align}\label{eq:A.5}
    n(\tilde{r},\tilde{\phi},\xi)
    &=
    n_+(\tilde{r},\tilde{\phi},\xi)+n_-(\tilde{r},\tilde{\phi},\xi) \nonumber \\
    &=
    \frac{1}{2}
    \frac{2-t}{\sqrt{1-t}}
    +
    g\tilde{r}
    \frac{4-3t}{4\sqrt{1-t}}
    \cos\tilde{\phi}
    \nonumber\\
    &=
    \frac{2\tilde{r}^2-\tilde{r}_b^2}{2\tilde{r}\sqrt{\tilde{r}^2-\tilde{r}_b^2}}
    +
    g
    \frac{4\tilde{r}^2-3\tilde{r}_b^2}{4\sqrt{\tilde{r}^2-\tilde{r}_b^2}}
    \cos\tilde{\phi}
    .
    \end{align}
As a consequence of the singularity in the shock wave, plasma density has a square root singularity at the boundary of the bubble.

We switch back to the initial coordinates $r,\phi,\xi$ and keep only terms of the order $g$: 
\begin{align}
\label{eq:pld_FF}
    &n(r,\phi,\xi)=\frac{2r^2-\tilde{r}_b^2}{2r\sqrt{r^2-\tilde{r}_b^2}} \nonumber \\ &+g \frac{\cos \phi}{4 \left(r^2-\tilde r_b^2 \right)^{3/2}}\left[\left(4 r^2-3 \tilde{r}_b^2 \right)\left(r^2-\tilde{r}_b^2 \right)-\frac{\tilde r_b^6}{8 r^2} \right].
\end{align}
As expected the first term in Eq.\eqref{eq:pld_FF} coincides with the electron density in the uniform case given in Ref.\cite{Stupakov:2016} while the second term provides correction that accounts for the initial transverse plasma gradient. 
%
\section{Pseudopotential \label{sec:PPT}}
%

Equation for the pseudopotential under the assumption of the non-relativistic plasma flow ($v_z\ll 1$) reads
\begin{align}
\label{eq:WPbl}
\Delta_\perp\psi=n_e-n_i,
\end{align}
 here $n_e$ is the plasma electron density and $n_i$ is the ion density that depends on $x$. In what follows, we will assume that
 \begin{align}
\label{eq:iond}
     n_i(x)
     =
     1+gx.
 \end{align}
We assume $r_0\ll 1$ and $r\ll 1$ switch to $\tilde x = x+g\nu\xi /2$, $\tilde y=y$ and $\tilde{r} = \sqrt{\tilde x^2+\tilde{y}^2}$ and the angle $\tilde{\phi} = \arccos{\tilde{x}/ \tilde{r}}$, account for the Eq.\eqref{eq:A.5} and Eq.\eqref{eq:iond} and rewrite Eq.\eqref{eq:WPbl} in the extended form
\begin{align}
\label{eq:WPexp}
\frac{1}{\tilde r}\frac{\partial}{\partial \tilde r}\left[\tilde r\frac{\partial \psi}{\partial \tilde r}\right]&+\frac{1}{\tilde r^2}\frac{\partial^2 \psi}{\partial \tilde\phi^2}=
\frac{2\tilde{r}^2-\tilde{r}_b^2}{2\tilde{r}\sqrt{\tilde{r}^2-\tilde{r}_b^2}}\theta(\tilde r -\tilde{r}_b)-1\\ \nonumber
    &+g\tilde{r} \cos \tilde \phi \left(
    \frac{4\tilde{r}^2-3\tilde{r}_b^2}{4\tilde{r}\sqrt{\tilde{r}^2-\tilde{r}_b^2}}
    \theta(\tilde r -\tilde{r}_b)-1\right) ,
\end{align}
Next we decompose $\psi$ in in a Fourier series
\begin{align}
\label{eq:psiFs}
    \psi(\tilde r,\tilde \phi)=\sum \tilde \psi_n (\tilde r) \cos(n \tilde \phi).
\end{align}
Consequently we have
\begin{align}
\label{eq:psi0}
\frac{1}{\tilde r} \frac{\partial}{\partial \tilde r}\left[\tilde r \frac{\partial \tilde \psi_0}{\partial \tilde r}\right]=\frac{2\tilde{r}^2-\tilde{r}_b^2}{2\tilde{r}\sqrt{\tilde{r}^2-\tilde{r}_b^2}}\theta(\tilde r -\tilde{r}_b)-1,
\end{align}
and 
\begin{align}
\label{eq:psi1}
\frac{1}{\tilde r} \frac{\partial}{\partial \tilde r}\left[\tilde r \frac{\partial \tilde \psi_1}{\partial \tilde r}\right]&-\frac{\tilde \psi_1}{\tilde r^2}= \\ \nonumber &g\tilde{r} \left(
    \frac{4\tilde{r}^2-3\tilde{r}_b^2}{4\tilde{r}\sqrt{\tilde{r}^2-\tilde{r}_b^2}}
    \theta(\tilde r -\tilde{r}_b)-1\right).
\end{align}
We introduce new normalized radius $\kappa=\tilde r/\tilde{r}_b$ and rewrite Eq.\eqref{eq:psi0} and  Eq.\eqref{eq:psi1} in the universal form
\begin{align}
\label{eq:psi0U}
\frac{1}{\kappa} \frac{\partial}{\partial \kappa}\left[\kappa \frac{\partial \tilde \psi_0}{\partial \kappa}\right]=\tilde{r}_b^2\left[\frac{2\kappa^2-1}{2\kappa\sqrt{\kappa^2-1}}\theta(\kappa -1)-1\right],
\end{align}
\begin{widetext}
and

\begin{align}
\label{eq:psi1U}
\frac{1}{\kappa} \frac{\partial}{\partial \kappa}\left[\kappa \frac{\partial \tilde \psi_1}{\partial \kappa}\right]-&\frac{\tilde \psi_1}{\kappa^2}= \tilde{r}_b^3 g\kappa \left[
    \frac{4\kappa^2-3}{4\kappa\sqrt{\kappa^2-1}}
    \theta(\kappa -1)-1\right].
\end{align}

Equations above could be integrated and the solutions read
\begin{align}
\label{eq:psi0f}
    \tilde \psi_0(\kappa)=\left \{ \begin{array}{c}
 a_1+a_2 \log\kappa-\frac{1}{4}\tilde{r}_b^2\kappa^2~~\kappa<1 \\
 a_3+a_4 \log\kappa+\frac{1}{4}\tilde{r}_b^2\kappa(\sqrt{\kappa^2-1}-\kappa)-\frac{1}{4}\tilde{r}_b^2 \log\left(\kappa+\sqrt{\kappa^2-1} \right)~~1<\kappa\ll\infty \\
\end{array} \right.,
\end{align}
and
\begin{align}
\label{eq:psi1f}
    \tilde \psi_1(\kappa)=\left \{ \begin{array}{c}
 b_1\kappa+\frac{b_2}{\kappa}-\frac{1}{8}g \tilde{r}_b^3\kappa^3~~\kappa<1 \\
 b_3\kappa+\frac{b_4}{\kappa}-\frac{1}{16} \tilde{r}_b^3 g \kappa \left\{2\kappa^2-2\kappa\sqrt{\kappa^2-1}+\log \left[\frac{\kappa+\sqrt{\kappa^2-1}}{\kappa-\sqrt{\kappa^2-1}}\right] \right\}~~1<\kappa\ll\infty \\
\end{array} \right. .
\end{align}
Here $a_i$ and $b_i$ are constants that could be found from the condition $\psi(r=0)<\infty$ for $\xi\neq 0$ and continuity of the pseudopotential and its derivative at the bubble boundary. 

First, we consider the monopole part $\tilde \psi_0 \left( \frac{\tilde r}{\tilde{r}_b}\right)$ Eq.\eqref{eq:psi0f}.
Condition $\tilde \psi_0(\tilde r=0)<\infty$ for $\xi\neq 0$ leads to $a_2=0$, continuity of the potential gives $a_3=a_1$ and continuity of the derivative requires $a_4=0$. Consequently we arrive at
\begin{align}
\label{eq:psi0m}
     \tilde \psi_0\left( \frac{\tilde r}{\tilde{r}_b}\right)=\left \{ \begin{array}{c}
 a_1-\frac{\tilde r^2}{4}~~\tilde r<\tilde{r}_b \\
 a_1+\frac{\tilde r}{4}(\sqrt{\tilde r^2-\tilde{r}_b^2}-\tilde r)-\frac{\tilde{r}_b^2}{4}\log\left(\frac{\tilde r}{\tilde{r}_b}+\sqrt{\left(\frac{\tilde r}{\tilde{r}_b}\right)^2-1} \right)~~\tilde{r}_b<\tilde r\ll\infty \\
\end{array} \right..
\end{align}

Next, we consider the dipole part $\tilde \psi_1 \left( \frac{\tilde r}{\tilde{r}_b}\right)$ Eq.\eqref{eq:psi1f}. Condition $\tilde \psi_1(\tilde r=0)<\infty$ for $\xi\neq 0$ leads to $b_2=0$ continuity of the potential gives $b_1=b_3+b_4$ and continuity of the derivative requires $b_1=b_3-b_4$. Consequently $b_4=0$, $b_3=b_1$ and we arrive at 
\begin{align}
\label{eq:psi1m}
    \tilde \psi_1 \left( \frac{\tilde r}{\tilde{r}_b}\right)= \left \{ \begin{array}{c}
 b_1\frac{\tilde r}{\tilde{r}_b}-\frac{1}{8}g\tilde r^3~~\tilde r<\tilde{r}_b \\
 b_1\frac{\tilde r}{\tilde{r}_b}-\frac{1}{16} g \tilde r \left\{2\tilde r^2-2\tilde r\sqrt{\tilde r^2-\tilde{r}_b^2}+\log \left[\frac{\tilde r+\sqrt{\tilde r^2-\tilde{r}_b}}{\tilde r-\sqrt{\tilde r^2-\tilde{r}_b}}\right] \right\}~~\tilde{r}_b<\tilde r\ll\infty \\
\end{array} \right..
\end{align}
\end{widetext}
We note that at large $\tilde{r} \gg \tilde{r}_b$ particular solution for the $\tilde \psi_0$ diverges as $\sim -\frac{\tilde{r}_b^2}{4}\log \tilde{r}$ and particular solution for the  $\tilde \psi_1$ diverges $\sim -\frac{\tilde{r}_b^2 g}{8}\tilde{r} \log \tilde{r}$ (See Appendix \ref{app:div} for the details). This is connected with the fact that we neglected screening effects in the considered approximation.

We switch back to the original coordinates $r$ and $\phi$. Noticing that $\cos \phi = \cos \tilde \phi +\mathcal{O}[g]$ and accounting for the fact that $\tilde \psi (\kappa) \sim \mathcal{O}[g]$ one may write
\begin{align}
\label{eq:ppfull}
    \psi(r,\phi,\xi)=&\tilde \psi_0 \left( \frac{r}{\tilde{r}_b}\right)+g\frac{\tilde{r}_b}{16}\tilde \psi^\prime_0 \left( \frac{r}{\tilde{r}_b}\right) \cos \phi \nonumber\\+ &\tilde \psi_1 \left(\frac{r}{\tilde{r}_b} \right) \cos \phi +\mathcal{O}[g^2].
\end{align}

We observe that, as expected, pseudopotential consists of two parts: monopole - that corresponds to the term $\tilde \psi_0 \left( \frac{r}{\tilde{r}_b}\right)$ and dipole - that is a combination of the total derivative by $\kappa$ of the monopole term $\tilde \psi^\prime_0 \left( \kappa \right)$ and a correction $\tilde \psi_1 \left(\frac{r}{\tilde{r}_b} \right) $. 

At distances, $r \sim 1$ plasma density should be unperturbed, and plasma electrons screen the field that arises from the bubble. This, in turn, results in the vanishing of the pseudopotential. To account for this and estimate the remaining unknown constants $a_1$ and $b_1$ we request pseudopotential to be zero at $r =1$. 

Fist, we notice that at $r \gg \tilde{r}_b$ expressions for the $\tilde \psi_0$ and $\tilde \psi_1$ reduces to
\begin{align}
    &\tilde \psi_0 \left( \frac{r}{\tilde{r}_b} \right) = a_1-\frac{\tilde{r}_b^2}{8} - \frac{\tilde{r}_b^2}{4} \log \left( 2 \frac{r}{\tilde{r}_b} \right)+\mathcal{O} \left[\frac{\tilde{r}_b}{r} \right], \\ \nonumber 
    &\tilde \psi_1 \left( \frac{r}{\tilde{r}_b} \right) = b_1 \frac{r}{\tilde{r}_b} -\frac{g \tilde{r}_b^2}{16}r- \frac{g \tilde{r}_b^2}{8}r \log \left( 2 \frac{r}{\tilde{r}_b} \right)+\mathcal{O} \left[\frac{\tilde{r}_b}{r} \right].
\end{align}
Next, keeping only divergent terms and setting $r=1$ we arrive at
\begin{align}
\label{eq:fin_const}
    a_1 \approx \frac{\tilde{r}_b^2}{4} \log \left(\frac{2}{\tilde{r}_b} \right), \nonumber \\
    b_1 \approx \frac{g \tilde{r}_b^3}{8} \log \left(\frac{2}{\tilde{r}_b} \right).
\end{align}

Following Eqs.\eqref{eq:psi0m} and \eqref{eq:psi1m} with Eq.\eqref{eq:ppfull} and Eq.\eqref{eq:fin_const} we arrive at the final expression for the pseudopotential inside the bubble in the form
\begin{align}
\label{eq:pspF}
    \psi(r,\phi,\xi)\approx &\frac{\tilde{r}_b^2}{4}\log \left(\frac{2}{\tilde{r}_b} \right)-\frac{r^2}{4}\nonumber\\&-g\frac{\tilde{r}_b^2 \cos \phi}{8} \left\{\frac{ r}{4}- r \log \left( \frac{2}{\tilde{r}_b}\right)+\frac{r^3}{\tilde{r}_b^2}  \right \}.
\end{align}
For the analysis it is convenient to normalize pseudopotential to $1/\tilde{r}_b^2$. We recall the definition of $\tilde{r}_b$ given by Eq.\eqref{eq:rb_sm} and write the expression for the normalised pseudopotential
\begin{align}
\label{eq:pspN}
   &\frac{\psi}{\tilde{r}^2_b}=-\frac{1}{8}\log \left(2\nu \xi \right)-\frac{x_n^2+y_n^2}{4}\nonumber\\&-g\frac{\sqrt{2\nu \xi}}{4} x_n\left\{\frac{1}{4}+ \frac{1}{2}\log \left(2\nu \xi \right)+x_n^2+y_n^2  \right \}, 
\end{align}
with $x_n=x/r_b$ and $y_n=y/r_b$.

\begin{figure}[t]
	\centering
	\includegraphics[width=1.\linewidth]{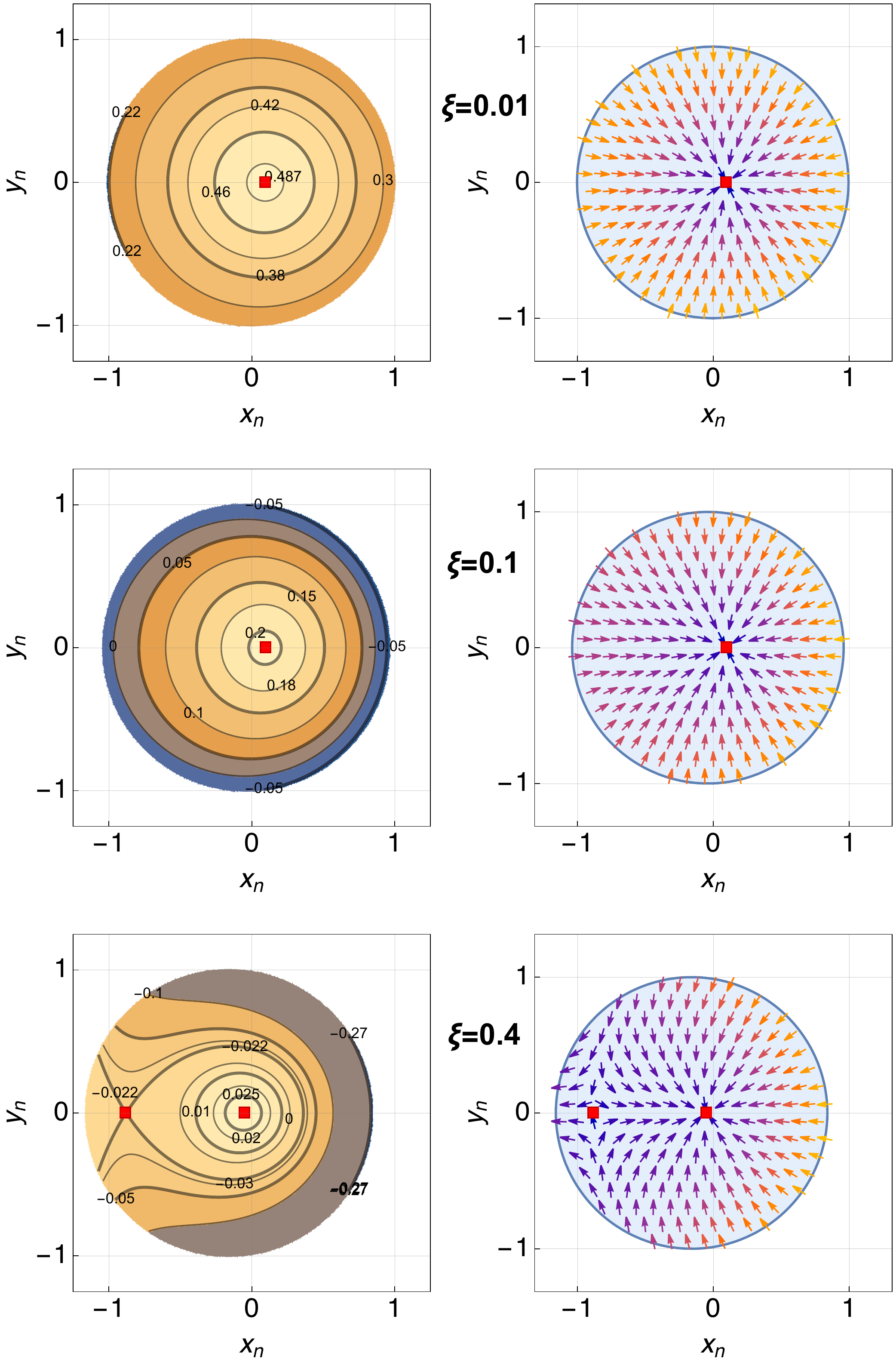}
	\caption{Contour plot for the normalized pseudopotential $\psi/\tilde{r}_b^2$ given by Eq.\eqref{eq:pspN} (left column) and normalized transverse wakefield $\mathbf{F}_\perp/\tilde{r}_b$ vector field given by Eq.\eqref{eq:trwN} (right column) for three different values of the longitudinal coordinate $\xi$ and $\nu=1$, $g=0.8$. Red dots indicate fixed points of the vector field given by Eq.\eqref{eq:fxpts}. We note that $x_n=x/\tilde{r}_b$ and $y_n=y/\tilde{r}_b$.\label{Fig:2}}
\end{figure}

We note that this equation is valid inside the bubble only when $r<r_b\ll1$, implying a small vicinity of the driver or a low charge regime. 

\section{Analysis}

With the help of the Eq.\eqref{eq:lor_main} and Eq.\eqref{eq:pspF} transverse part of the Lorentz force per unit charge of the negatively charged test particle could be evaluated as
\begin{align}
   &F_x=-\frac{x}{2}-g\frac{\tilde{r}_b^2 }{8}\left[\frac{1}{4}-\log \left(\frac{2}{\tilde{r}_b} \right)+\frac{3x^2+y^2}{\tilde{r}_b^2} \right], \nonumber \\ 
   &F_y=-\frac{y}{2}-g\frac{xy}{4}.
\end{align}
It is convenient to normalize it to $1/\tilde{r}_b$ and present in terms of $x_n=x/\tilde{r}_b$ and $y_n=y/\tilde{r}_b$
\begin{align}
\label{eq:trwN}
   &\frac{F_x}{\tilde{r}_b}=-\frac{x_n}{2}-g\frac{\sqrt{2 \nu\xi} }{4}\left[\frac{1}{4}+\frac{1}{2}\log \left(2\nu \xi \right)+3x_n^2+y_n^2 \right], \nonumber \\ 
   &\frac{F_y}{\tilde{r}_b}=-\frac{y_n}{2}-g\sqrt{2 \nu \xi}\frac{x_n y_n}{2}.
\end{align}
We note that pseudopotential has a cubic term in $x_n$ that naturally leads to two fixed points of the vector field (one stable and one unstable). 
By setting $F_x=0$ and $F_y=0$ one may find fixed points of the transverse wakefield by solving the corresponding algebraic system that follows from Eq.\eqref{eq:trwN}: 
\begin{align}
\label{eq:fxpts}
   &y_n^F=0, \nonumber \\
   &x_n^{S}=-\frac{g\sqrt{\nu \xi}}{4\sqrt{2}}\left[1+2\log \left(2 \nu \xi \right) \right], \\
   &x_n^{uS}=-\frac{\sqrt{2}}{3g \sqrt{\nu \xi}}+\frac{g\sqrt{\nu \xi}}{4\sqrt{2}}\left[1+2\log\left(2 \nu \xi \right) \right]. \nonumber
\end{align}
To simplify the final formula, we kept only terms of the order $g$, i.e. we disregarded terms of the order $\mathcal{O}\left[g^2 \right]$ in Taylor decomposition.

 In Fig.\ref{Fig:2} we show level sets of the pseudopotential given by Eq.\eqref{eq:pspN}, transverse wakefield vector field Eq.\eqref{eq:trwN} and fixed points of the vector field Eq.\eqref{eq:fxpts} for three different values of the longitudinal coordinate $\xi$. We chose extreme (and probably unreachable in practice) parameters of $g=0.8$ and $\nu=1$ to emphasize the effect. We observe that at small values of $\xi$, where the model is directly applicable, only one stable fixed point exists within the bubble cross-section. Transverse gradient shifts the electromagnetic origin towards the higher densities of the ion column, but the net effect remains focusing albeit asymmetric. Further increase in $\xi$ does not change the picture. The asymmetry in the focusing grows, but the structure of the wake remains the same. Interestingly, if we speculate and go beyond the formal applicability of the considered model. We may observe the situation when both stable and unstable fixed points are located inside the bubble cross-section. We point out, that despite the complex structure of the pseudopotential, the stable region (the region where the beam is attracted to the stable fixed point) occupies more than half of the bubble cross-section even in this unrealistic scenario. The latter indicates that most likely fixed transverse plasma gradient (a transverse plasma gradient that does not change in $z$) should not affect the driver dynamics (at least within considered approximation) and only results in some asymmetric distortion of the bubble shape and wake.
 
It is worth mentioning that the plasma gradient may fluctuate randomly due to the random fluctuations of the plasma density. Such random fluctuation will result in a random kick. It is well known (see Refs.\cite{RandomF1,Delayen}) that random kicks may lead to emittance growth and potentially may lead to driver instability. Indeed, in 1D emittance growth (see for instance Ref.\cite{RandomF2}) due to the random kick reads
\begin{align}
    \frac{\delta \varepsilon}{\delta s}=\frac{\langle x^2\rangle \langle F_x^2\rangle}{\varepsilon}.
\end{align}
Here $\varepsilon$ is the beam emittance and $\tau$ is the characteristic time of the fluctuation. Following Eq.\eqref{eq:trwN} we can write
\begin{align}
\langle F_x^2\rangle\sim \langle g^2\rangle \sim \langle n^2\rangle-n_0^2.    
\end{align}
Consequently, the dispersion of the density fluctuation sets the growth rate for the emittance. This observation motivates further studies in more realistic scenarios by either applying Lu model \cite{Lu} or proper extension of a numerical simulation \cite{QPIC,warp,OSIRIS,EPOCH,PICon,FBPIC}.   

\section{Conclusions}

We have presented a detailed analysis of the wakefield in the presence of the transverse plasma gradient. A simple ballistic model from Refs.\cite{Stupakov:2016,Stupakov:2018} was updated to account for the linear transverse inhomogeneity in plasma. As a result, we provide final analytic expressions for the pseudopotential and the transverse wakefield. We note that as in the flat bubble regime considered previously in Ref.\cite{FlatPRAB}, the bubble shape in the present study shows similar distortion. Namely, a small perturbation to the plasma density results in the “bending” of the bubble toward a lower plasma gradient. However, in contrast to the flat bubble regime, in the round bubble transverse wake does not vanish.  

We point out that random fluctuation of the plasma density, which naturally occurs, may lead to emittance growth and potentially become a challenge. Consequently, further developments in this direction are in order. 

We note that the numerical examples provided in the paper
are synthetic, and we choose parameters for these examples to emphasize corresponding effects. In reality, the parameter
g— a transverse plasma gradient should be on the order of
1$\%$ or less as well as $\zeta\ll 1$. We emphasize that the whole analysis is applicable only when plasma electrons are nonrelativistic. 

Despite the restrictions outlined above, the model presented is still useful, as it is complementary to the Lu model of the plasma bubble
\cite{Lu}. The model presented could be "merged" with the Lu model such that the results of the ballistic model may serve as
an initial condition for the Lu equation. A combined model will be free of the empiric parameters and cover the whole range of the driver beam intensities. 

The equation derived in the present paper could be used as a crude estimate for the transverse emittance growth due to the random fluctuations of the plasma density.  

\appendix

\section{Envelope surfaces for the ballistic trajectories \label{app:envs}}

First, we notice that if $g=0$, then as was shown in Ref.~\cite{Stupakov:2016}, for a given $\xi$, the map~Eq.\eqref{eq:trjf} when $r_0$ varies from 0 to $\infty$ and $\phi_0$ varies from 0 to $2\pi$ leaves and empty circle of radius $2\sqrt{2\nu\xi}$ centered at $x=y=0$. With a nonzero $g$, we can move the term $-\nu\xi g/2$ from the right to the left-hand side. We then see that this empty circle is shifted by $-\nu\xi g/2$ along $x$, and hence its equation is Eq.~\eqref{eq:103}.


Another approach is to consider an arbitrary ballistic trajectory as given by Eq.\eqref{eq:trjf}. This trajectory could be represented in a vector form in $xy\xi$ space as
\begin{align}
\label{eq:trvec}
    r=\left(x,y,\xi \right)^{\mathrm{T}}
\end{align}
with $x$ and $y$ given by Eq.\eqref{eq:trjf}. 
Next we consider a transformation of the $xy\xi$ space along $\xi$ axis given by a rotation matrix 
\begin{align}
\label{eq:Rxi}
    R_\xi=\left(
\begin{array}{ccc}
 \cos \phi_0 & \sin \phi_0  & 0 \\
 -\sin \phi_0 & \cos \phi_0  & 0 \\
 0 & 0 & 1 \\
\end{array}
\right)
\end{align}
and apply to Eq.\eqref{eq:trvec}. With Eq.\eqref{eq:trjf} we have
\begin{align}
\label{eq:r1}
    &R_\xi r = \nonumber \\ &\left( r_0+\frac{2\xi \nu}{r_0}-\frac{\xi g \nu}{2} \cos\phi_0,\frac{\xi g \nu}{2} \sin\phi_0,\xi \right)^{\mathrm{T}}. 
\end{align}
Next we consider rotation along $x$ axis
\begin{align}
\label{eq:Rx}
   R_x=\left(
\begin{array}{ccc}
 1 & 0 & 0 \\
 0 & \cos \theta  & -\sin \theta  \\
 0 & \sin \theta & \cos \theta \\
\end{array}
\right)
\end{align}
such that $\tan(\theta)=\frac{\nu g}{2} \sin (\phi_0)$. Combining Eq.\eqref{eq:Rx} with Eq.\eqref{eq:r1} we get
\begin{align}
\label{eq:transtrj}
       &R_xR_\xi r = \\ \nonumber &\left( r_0+\frac{2\xi \nu}{r_0}-\frac{\xi g \nu}{2} \cos\phi_0,0,\xi \sqrt{ 1+\left[\frac{g \nu}{2}\sin\phi_0\right]^2} \right)^{\mathrm{T}}. 
\end{align}
It follows from Eq.\eqref{eq:transtrj} that $xy\xi$ space could be rotated with the help of $R_xR_\xi$ transformation such that after the combiner rotation any given trajectory will always lay in the $Ox\xi$ plane. Combiner rotation $R_xR_\xi$ depends only on the initial polar angle $\phi_0$ of the trajectory starting point and is independent of $r_0$. Consequently, we conclude that all trajectories with starting points with the same initial angle $\phi_0$ are transformed by the rotation $R_xR_\xi$ to the $Ox\xi$ plane as well. It is worth reiterating that all trajectories that start at the same initial angle $\phi_0$ will stay in the same plane and, with the help of two rotations $R_\xi$ and $R_x$, could be always translated into $Ox\xi$ plane. As far as the transformation $R_x R_\xi$ is not degenerate and has only one stationary point $(0,0,0)$ trajectories that start from different angles $\phi_0$ never cross as different $\phi_0$ defines different rotations of the $xy\xi$ space. This, in turn, effectively reduces the initial problem of finding an envelope surface for all trajectories to a problem of finding an envelope curve for families of the trajectories with the same angle $\phi_0$ transformed to the $Ox\xi$ plane with the help of $R_xR_\xi$ rotation.

If $\tilde{x}\equiv r_0+\frac{2\xi \nu}{r_0}-\frac{\xi g \nu}{2} \cos(\phi_0)$ then an envelope curve for each family according to Eq.\eqref{eq:transtrj} could be written as
\begin{align}
    \tilde{x}_e=2\sqrt{2\nu \xi}-\frac{\xi g \nu}{2}\cos(\phi_0).
\end{align}
Thus, points on an envelope curve in a transformed plane have the coordinates 
\begin{align}
\label{eq:transtrje}
      &\tilde{r}_e = \\ \nonumber &\left( 2\sqrt{2\nu \xi}-\frac{\xi g \nu}{2}\cos\phi_0,0,\xi \sqrt{ 1+\left[\frac{g \nu}{2}\sin\phi_0\right]^2} \right)^{\mathrm{T}}. 
\end{align}
Inverse transformation $R^{-1}_\xi R_x^{-1}$ applied to the Eq.\eqref{eq:transtrje} gives a set of points that resemble envelope curve that results from the trajectories that have fixed polar angle $\phi_0$ for the starting points.
\begin{align}
    &R^{-1}_\xi R_x^{-1}\tilde{r}_e= \\ \nonumber &\left( 2\sqrt{2\nu \xi} \cos\phi_0-\frac{\xi g \nu}{2},2\sqrt{2\nu \xi} \sin\phi_0,\xi \right)^{\mathrm{T}}.
\end{align}
If now $x\equiv 2\sqrt{2\nu \xi} \cos\phi_0-\frac{\xi g \nu}{2}$ and $y\equiv2\sqrt{2\nu \xi} \sin\phi_0$ then one can arrive at the Eq.\eqref{eq:103}.

The analysis presented above has another important consequence. As far as each $\phi_0$ family of the trajectories forms a separate and independent of others envelope line, plasma density that results from electron blowout could be derived exactly the same way for each family in the transformed plane.

\section{Divergence of the unscreened pseudopotential for the large values of  \texorpdfstring{$\kappa$}{kappa} \label{app:div}}

We consider an asymptotic at $\kappa\gg1$ of the right hand side of the Eq.\eqref{eq:psi1U} 
\begin{align}
   r_b^3 g\kappa &\left[
    \frac{4\kappa^2-3}{4\kappa\sqrt{\kappa^2-1}}
    \theta(\kappa -1)-1\right]\approx \nonumber\\ &r_b^3 g\kappa \left[
    \theta(\kappa -1)-1-\frac{\theta(\kappa-1)}{4\kappa^2}\right]. 
\end{align}
With this for $\kappa>1$ Eq.\eqref{eq:psi1U} reduces to
\begin{align}
\label{eq:psi1Ut2}
\frac{1}{\kappa} \frac{\partial}{\partial \kappa}\left[\kappa \frac{\partial \tilde \psi_1}{\partial \kappa}\right]-\frac{\tilde \psi_1}{\kappa^2}\approx-\frac{r_b^3 g}{4\kappa}.
\end{align}
Particular solution has the form
\begin{align}
     \tilde \psi_1\approx \frac{r_b^3 g\kappa}{16}\left(1-2\log \kappa\right)
\end{align}
accounting for the $\log \kappa\gg1$ we get
\begin{align}
    \tilde \psi_1\approx -\frac{r_b^3 g}{8}\kappa \log \kappa.
\end{align}

\begin{acknowledgments}
The author is grateful to G. Stupakov for fruitful discussions.The work was supported by the Foundation for the Advancement of Theoretical Physics and Mathematics "BASIS'' $\#$22-1-2-47-17 and ITMO Fellowship and Professorship program.
\end{acknowledgments}
\bibliographystyle{unsrt}
\bibliography{inhom_plasma}

\end{document}